%% file: root.tex
\pdfoutput=1
\documentclass[letterpaper, 10 pt, conference]{ieeeconf}  % Comment this line out if you need a4paper

\IEEEoverridecommandlockouts                              % This command is only needed if 
\usepackage[font=footnotesize]{caption} 
\renewcommand{\baselinestretch}{.97}

\usepackage{amsmath} % assumes amsmath package installed
\allowdisplaybreaks
\usepackage{amssymb}  % assumes amsmath package installed

\usepackage{algorithm}
\usepackage{algpseudocode}
\usepackage[textsize=small, disable]{todonotes}
\usepackage{siunitx}
\usepackage{tikz}
\usetikzlibrary{arrows, shapes, automata, matrix}
\usetikzlibrary{arrows.meta,decorations.markings}
\usepackage{tikzscale}
\usepackage{multirow}
\usepackage{rotating}
\usepackage{tabularx}
\usepackage{lipsum}
\usepackage{graphicx}
\usepackage{epstopdf}

\newtheorem{definition}{Definition}

\title{\LARGE \bf
On the Selection of Calculable Residual Generators for UAV Fault Diagnosis
}

\author{Georgios Zogopoulos Papaliakos and Kostas J. Kyriakopoulos% <-this % stops a space
\thanks{The authors are with Control Systems Lab, School of Mechanical Engineering, National Technical University of Athens, Greece
	{\tt\small \{gzogop, kkyria\}@mail.ntua.gr}}% <-this % stops a space
}

\begin{document}

\setlength{\abovedisplayskip}{3.8pt}
\setlength{\belowdisplayskip}{3.8pt}

\maketitle
\thispagestyle{empty}
\pagestyle{empty}

%%%%%%%%%%%%%%%%%%%%%%%%%%%%%%%%%%%%%%%%%%%%%%%%%%%%%%%%%%%%%%%%%%%%%%%%%%%%%%%%
\begin{abstract}

 Structural Analysis is an established method for Fault Detection and Identification (FDI) in large-scale systems, enabling the discovery of Analytical Redundancy Relations (ARRs) which serve as residual generators. However, most techniques used to enumerate ARRs do not specify the matching used to calculate each of those ARRs. This can result in non-implementable or unusable residual generators, in the presence of non-invertibilities in the equations involved or in lack of computational tools. In this paper, we propose a methodology which combines a priori and a posteriori information in order to reduce the time required to find implementable, usable residual generators of minimum cost. The method is applied to a fixed-wing Unmanned Aerial Vehicle (UAV) model.

\end{abstract}

%%%%%%%%%%%%%%%%%%%%%%%%%%%%%%%%%%%%%%%%%%%%%%%%%%%%%%%%%%%%%%%%%%%%%%%%%%%%%%%%
\input{introduction.tex}
\input{structural_analysis.tex}
\input{causality_calculability.tex}
\input{weights.tex}
\input{proposal.tex}
\input{example.tex}

\input{conclusion.tex}

%\input{appendix.tex}

%%%%%%%%%%%%%%%%%%%%%%%%%%%%%%%%%%%%%%%%%%%%%%%%%%%%%%%%%%%%%%%%%%%%%%%%%%%%%%%%

%\addtolength{\textheight}{-12cm}   % This command serves to balance the column lengths
                                  % on the last page of the document manually. It shortens
                                  % the textheight of the last page by a suitable amount.
                                  % This command does not take effect until the next page
                                  % so it should come on the page before the last. Make
                                  % sure that you do not shorten the textheight too much.

%%%%%%%%%%%%%%%%%%%%%%%%%%%%%%%%%%%%%%%%%%%%%%%%%%%%%%%%%%%%%%%%%%%%%%%%%%%%%%%%

%%%%%%%%%%%%%%%%%%%%%%%%%%%%%%%%%%%%%%%%%%%%%%%%%%%%%%%%%%%%%%%%%%%%%%%%%%%%%%%%

\bibliographystyle{IEEEtran}
\bibliography{myBibliograpy}

\end{document}

%% file: introduction.tex
\section{INTRODUCTION}

Even though there exist well-established algorithms for the control of commercial Unmanned Aerial Vehicles (UAVs), they are rarely designed to handle faults. Notably, faults are most common in commercial UAVs, since such vehicles are manufactured to lower quality standards, compared to military and aerospace applications.

A fault is a deviation of the system parameters or the system structure from the nominal condition. Most of the times, if faults are allowed to go unchecked, they will eventually result to failures, rendering the system inoperable. Hence, it is important to detect faults as early as possible and take counter-measures against them, in order to maintain system health and ensure its operational status.

It is reasonable to expect from modern UAVs to exhibit fault-tolerance to a certain degree and, to that goal, Fault Detection is indispensable. Fault Detection involves setting up mathematical and physical structures that are able to detect when a fault occurs in a system. Afterwards, Fault Isolation can be performed, in order to identify the system component which is under fault. This is an important piece of information, since it allows us to locate the source and nature of a fault \cite{Blanke2006a}.

In consistency-based diagnosis for continuous-time systems, the difference $r(t)=y(t)-\hat{y}(t)$ between a measured quantity $y(t)$ and a quantity estimate based on the nominal system behaviour $\hat{y}(t)$ is used to perceive faults. Under no-fault conditions, $r(t)=0$ should hold and vice versa. Such difference functions are called \textit{residuals}. Ultimately, a residual can be formulated as $r(t)=f(y(t))$, i.e. as a function of continuous-time measurements. $f$ is called a \textit{residual generator} function. The required information to perform FDI can be provided either by hardware redundancy or analytical redundancy. The first approach places multiple, redundant sensory equipment on the aircraft which contribute in validity checks. However, this is not a lucrative option in small UAVs, where low weight, cost and power consumption are desired features. Thus, the second approach, where redundant information is introduced by mathematical models, stands out as a prominent field of research. Finding Analytical Redundancy Relations (ARRs) which can be used as residual generators is a primary goal \cite{Blanke2006a, Fravolini2009, Krysander2008}.

In order to perform model-based FDI at the component level, it is necessary to enumerate and take into account the equations which make up the mathematical model of the UAV down to that level. However, such detailed representations will yield hundreds of equations, rendering the process of finding consistency checks very hard to do by hand. Formal, scalable methods are required to achieve this. Structural Analysis (SA) provides such methods \cite{Blanke2006a, Izadi-Zamanabadi2002} and has been used in linear systems \cite{Boukhobza2013}. Still, SA is not yet able to extract consistency checks for generic, non-linear systems automatically as well as in real-time, a prerequisite for UAS with high autonomy and reconfigurable controllers.

In this paper, we propose a methodology reducing the time needed to perform the residual selection task, by using a weighted bipartite graph structure and combining information available both a priori and a posteriori of the analysis.

In section \ref{sec: sa}, SA is presented in the context of residual generator extraction. In section \ref{sec: cc}, limitations in current SA techniques are mentioned. In section \ref{sec: we}, we introduce the concept of a weighted structural graph. In section \ref{sec: pr}, our proposed methodology is expanded upon. In section \ref{sec: ex}, an application example on a fixed-wing UAV model is presented. Finally, section \ref{sec: conclusion} concludes this work.

%% file: structural_analysis.tex
\section[sa]{STRUCTURAL ANALYSIS} \label{sec: sa}

SA is a qualitative modeling methodology which, given a mathematical model, captures only whether there exist relations between equations (also referred to as constraints) and variables. The resulting structure, although abstract, provides information on component interaction and facilitates FDI methods. The structure can be represented as\ a bipartite graph and graph-theoretic methods can be used to extract results in a methodical manner \cite{Blanke2006a},\cite{Izadi-Zamanabadi2002}.

%\subsection{A Sample UAV Model}
%A small system is provided to aid in the presentation of SA methodologies in the rest of this paper. It is a linearized, longitudinal model of a fixed-wing UAV, as presented in \cite{Beard2012}, modified for the needs of the analysis.
%\begin{IEEEeqnarray}{rCl}
%	\dot{u} & = & X_u u {+} X_w w {+} X_q q {-}g\cos(\theta ^*)\theta {+} X_{\delta_e} \delta_e {+} X_{\delta_t} \delta_t \label{eq: systemStart}\\
%	\dot{w} & = & Z_u u {+} Z_w w {+} Z_q q {-}g\cos(\theta ^*)\theta {+} Z_{\delta_e} \delta_e \label{eq: solveExample} \\
%	\dot{q} & = & M_u u {+} M_w w {+} M_q q {+}  M_{\delta_e} \delta_e\\
%	\dot{\theta} & = & q\\
%	\dot{h} & = & \sin(\theta ^*) u {-}\cos(\theta ^*) w  {+} ( u^*\cos(\theta ^*) {+}   w^*\sin(\theta ^*)) \theta 	
%\end{IEEEeqnarray}
%The state vector of this model is $\mathbf{X}{=}\left[u,w,q,\theta,h\right]^\intercal$, representing x-axis body-frame velocity, z-axis body-frame velocity, pitch rate, pitch angle and altitude respectively. $\theta^*, u^*$ and $w^*$ are known linearization points. $\mathbf{U} {=} \left[ \delta_e, \delta_t \right]^\intercal$ is the known input variable vector.
%
%Additionally, we wish to add information on ground elevation, $h_{gr}$, and the ground altitude of the UAV, $d_{gr}$.
%\begin{equation}
%	d_{gr} =  h - h_{gr} \label{eq: extraEq}
%\end{equation}
%Assuming the UAV is equipped with sensors to measure $q$ and $\theta$, we add the following equations to our model.
%\begin{IEEEeqnarray}{rCl}
%	y_1 & = & q \label{eq: qsensor}\\
%	y_2 & = & \theta \label{eq: systemEnd}
%\end{IEEEeqnarray}

\subsection{The Structural Model}

We construct the set of model constraints $\mathcal{C}$, with elements $c_i$, corresponding to each of the equations comprising the mathematical model of our system. In this work, we choose to expand $\mathcal{C}$ by adding the differentiation constraints separately for all differentiated variables, instead of inserting differentiated versions of existing constraints; we treat a variable and its derivative as structurally different \cite{Blanke2006a}.
%\begin{IEEEeqnarray}{rCl}
%	\dot{u} & = & (d/dt)u \label{eq: derStart} \\
%	\dot{w} & = & (d/dt)w \\
%	\dot{q} & = & (d/dt)q \\
%	\dot{\theta} & = & (d/dt)\theta \\
%	\dot{h} & = & (d/dt)h \label{eq: derEnd}
%\end{IEEEeqnarray} 
Given the set of available model constraints $\mathcal{C}$, we construct the structural model of the system in the form of an undirected bipartite graph $ \mathbf{G} {=} \left(\mathcal{C},\mathcal{X},\mathcal{E}\right)$ \cite{Blanke2006a, Izadi-Zamanabadi2002, Krysander2005}.
$\mathcal{X}{{=}}\mathcal{X}_U \bigcup \mathcal{X}_K$ is the set of involved variables, composed by the set of known variables $\mathcal{X}_K$ (such as inputs and measurements) and the set of unknown variables $\mathcal{X}_U$.
We can write that
$\mathcal{X}{{=}}var(\mathcal{C})$, 
$ \mathcal{X}_K{=}var_K(\mathcal{C})$ and 
$\mathcal{X}_U {=} var_U(\mathcal{C})$ to denote all, known and unknown variables in $\mathcal{C}$.
$\mathcal{E}$ is the set of edges connecting $\mathcal{C}$ and $\mathcal{X}$, $\mathcal{E}{=}\left\{ e_{ij}{=}(c_i, x_j) : x_j \in var(c_i) \right\}$.

%A visualization of the resulting bipartite graph can be seen in Fig. \ref{fig: ex_sm}. Constraints are depicted as bars, labeled for clarity: System equations \ref{eq: systemStart}-\ref{eq: extraEq} are labeled $e_{1}$-$e_6$, measurement equations \ref{eq: qsensor}-\ref{eq: systemEnd} are labeled $m_1$,$m_2$ and differential equations \ref{eq: derStart}-\ref{eq: derEnd} are labeled $d_1$-$d_5$. Known variables are depicted as gray circles and unknown variables as white circles. Known variables could have been omitted from the structural representation without any loss of information, as far as fault diagnosis is concerned.
%\begin{figure}
%	\centering
%	\resizebox{.75\linewidth}{!}{
%		\includegraphics[]{figures/ex_sm.tikz}
%	}
%%	\input{figures/ex_sm.tex}
%	\caption{The structural model of the sample system}
%	\label{fig: ex_sm}
%\end{figure}

A bipartite graph can also be represented as a biadjacency matrix, whose rows correspond to constraints and columns to variables. The biadjacency matrix $\mathbf{A}$ is defined as
\begin{equation}
a_{ij} = 
\begin{cases}
	1 & \text{iff } (c_i, x_j)\in\mathcal{E}, x_j \in \mathcal{X} \\
	0 & \text{iff } (c_i, x_j)\notin\mathcal{E}, x_j \in \mathcal{X}
\end{cases}
\end{equation}
Without loss of structural information we can use only $\mathcal{X}_U$ to populate the columns.
%The adjacency matrix of our example system can be seen in Fig. \ref{fig: ex_am}.
%\begin{figure}
%\centering
%\resizebox{0.75\linewidth}{!}{
%	\input{figures/ex_am.tex}
%}
%\caption{The biadjacency matrix of the example system}
%\label{fig: ex_am}
%\end{figure}
%In a best-case situation, if it holds that $|\mathcal{C}|{>}|\mathcal{X}_U|$, then there is analytical redundancy in the graph and it may be possible to extract ARRs from it.

\subsection{Solving the System Graph}

While searching for ARRs, it is of interest to solve as many equations for one of their variables as possible. Structurally, this is equivalent to assigning each variable vertex to one constraint vertex, respecting the edge set. The notion of \textit{matching} of a graph is very useful for that purpose.

Matching is a subset of $\mathcal{E}$ such that $\mathcal{M} {=} \left\{ m_i{=}(c_i, x_i) \in \mathcal{E} \right. $ $\left. : m_i {\neq} m_j \text{ iff } c_i {\neq} c_j \wedge x_i {\neq} x_j, \forall i,j\right\}$ \cite{Murota2000}, i.e. given the set of edges $\mathcal{E}$, we select edges so that no two edges have a common vertex. If the cardinality of $\mathcal{M}$ is the maximum possible for a given $\mathcal{E}$, then the matching is called \textit{maximal}.
If  $ \left| \mathcal{M} \right|{=}\left|\mathcal{X}_U\right|$ or $ \left| \mathcal{M} \right|{=}\left|\mathcal{C}\right|$, then the matching is called \textit{complete} with respect to  $\mathcal{X}_U$ or to $\mathcal{C}$ respectively. If $\left| \mathcal{M}\right|{=}\mathcal{X}_U{=}\mathcal{C}$ then the matching is \textit{perfect}.
For a given graph, the maximal matching may not be unique.
Unless the matching is complete on  $\mathcal{X}_U$, then no variable in  $\mathcal{X}_U$ is guaranteed to be calculable.

%Each matching proposes a pairing for variables of $\mathcal{X}_U$ to be solved by equations of $\mathcal{C}$, ensuring that each equation will be used only once. Matchings which are complete on  $\mathcal{X}_U$ allow for the calculation of all the unknown variables, in a structural sense. However, if the matching is not complete on $\mathcal{X}_U$, then not all variables covered by the matching are guaranteed to be calculable.

%Fig. \ref{fig: ex_matching} displays a possible matching of our sample graph. Matched edges (belonging in $\mathcal{M}$) are embossed.
%There is no point in including any variable from $\mathcal{X}_K$ in the matching, since they are already known.
%\begin{figure}
%	\centering
%	\resizebox{0.75\linewidth}{!}{
%		\input{figures/ex_match.tex}
%	}
%	\caption{A maximal matching of the sample system}
%	\label{fig: ex_matching}
%\end{figure}

After a matching $\mathcal{M}$ has been found, we can create a \textit{directed} graph $\mathbf{G}_d{=}\left\{\mathcal{C},\mathcal{X},\mathcal{E}_d \right\}$, with its edges defined as
$\mathcal{E}_d {=} \{e_{i,j}{=}(c_i, x_j) $$:$$ e_{i,j} {\in} \mathcal{M} \} $ $\bigcup$ $ \{ e_{i,j}{=}(c_i,x_j) $$:$$ (x_j, c_i) {\notin} \mathcal{M} \},$ $ e_{i,j} {\in} \mathcal{E}$.
The reverse of this graph is obtained by reversing the directionality of its edges and is denoted as $\mathbf{G}_d'$.

\begin{definition}[Reachability]
	Given an undirected (directed) graph $\mathbf{G}{=}\left\{\mathcal{C},\mathcal{X},\mathcal{E} \right\}$ and two vertices $v_i, v_j \in \mathcal{V} {=} \mathcal{X} {\bigcup} \mathcal{C}$, $v_j$ is \textit{reachable} from $v_i$ iff there is an undirected (directed) path in $\mathbf{G}$ from $v_i$ to $v_j$.
	A subset $\mathcal{V}_j \subseteq \mathcal{V}$ is reachable from $v_i$ iff there is an undirected (directed) path from $v_i$ to $v_j, \forall v_j {\in} \mathcal{V}_j$. A subset $\mathcal{V}_j$ is reachable from subset $\mathcal{V}_i$ iff there is an undirected (directed) path from $v_i$ to $v_j$, $\forall v_i {\in} \mathcal{V}_i, v_j {\in} \mathcal{V}_j$.
\end{definition}

\begin{figure}
	\centering
	\includegraphics[width=0.7\linewidth]{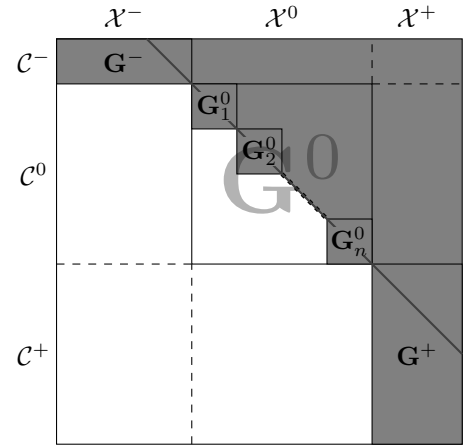}
	\caption{A typical Dulmage-Mendelsohn decomposition}
	\label{fig: dm}
\end{figure}

For any given bipartite graph, a unique decomposition on $\mathbf{A}$ is defined, called the Dulmage-Mendelsohn (DM) decomposition. It identifies three (possibly empty) subgraph components:
$\mathbf{G}^-{=}\left(\mathcal{C^-},\mathcal{X^-},\mathcal{E^-}\right)$, 
$\mathbf{G}^0{=}\left(\mathcal{C}^0,\mathcal{X}^0,\mathcal{E}^0\right)$ and
$\mathbf{G}^+{=}\left(\mathcal{C}^+,\mathcal{X}^+,\mathcal{E}^+\right)$.
The decomposition guarantees that there exists a complete matching on $\mathcal{C^-}$ in $\mathbf{G}^-$, a perfect matching in $\mathbf{G}^0$ and a complete matching on $\mathcal{X^+}$ in $\mathbf{G}^+$. Thus, $\mathbf{G}^-$ is called the \textit{under-constrained} part of $\mathbf{G}$, $\mathbf{G}^0$ \textit{just-constrained} and $\mathbf{G}^+$ \textit{over-constrained} \cite{Dulmage1958}. The decomposition can take a block-triangular form (Figure \ref{fig: dm}). The diagonal line represents the matchings between $\mathcal{C}$ and $\mathcal{X}$.
Since fault detection uses analytical redundancy in the form of unmatched constraints, which exist only in $\mathbf{G^+}$ it is evident that the only faults that can be detected are those which affect constraints belonging in $\mathbf{G^+}$.
\vspace{-2.5pt}

\subsection{Matching Algorithms}
%The ability to generate a matching is instrumental for all structural analysis methods.
There are several matching algorithms for undirected bipartite graphs, but two of them are the most common in SA applications.

The first one is the \textit{Ranking} algorithm, which was presented in \cite{Blanke2006a} and propagates an "information front" based on known variables. It has the property that every constraint it matches has exactly one variable unknown at that time, which is then solved for. The complexity of this algorithm is O($|C||X_U|$) \cite{Blanke2006}.
It is fast and can easily produce an acyclic calculation order, but cannot guarantee a complete matching when applied on cyclic graphs.

Another, more classical algorithm is the \textit{Hungarian method}.
The original algorithm \cite{Kuhn1955} and its multiple improvements \cite{Leiserson2009} have been a standard tool in graph theory and specifically for calculating maximum matchings in undirected bipartite graphs. They have low, polynomial computational cost, and are applicable for weighted undirectional graphs as well.
They can extract maximal matchings for cyclic graphs but $\mathcal{G}_d$ may contain cycles.

\subsection{MSOs}
Each unmatched constraint $c_i$ in $\mathbf{G}$ can be used as residual generator to detect faults in equations reachable from $c_i$, in $\mathbf{G}_d'$. In order to minimize the fault candidates for each triggered residual, it is of interest to find residuals which are sensitive to as few faults as possible.

The notion of Minimal Structurally Overdetermined sets (MSOs) is useful for that purpose \cite{Krysander2005}. An MSO is a set of equations $\mathcal{C}' \subseteq \mathcal{C}$ whose corresponding graph $\mathbf{G}' {=} \left\{\mathcal{C}', \mathcal{X}', \mathcal{E}' \right\}$ has the following properties:
\begin{enumerate}
	\item $\mathcal{X}' = var(\mathcal{C}')$
	\item $\mathcal{E}' = \left\{ (c_i, x_k) \in \mathcal{E} : c_i \in \mathcal{C}', x_k \in \mathcal{X}' \right\}$
	\item $G' = G'^+$
	\item $|\mathcal{C}'| = |\mathcal{X}_U'| + 1$
\end{enumerate}

With a slight abuse of notation, we will denote the corresponding graph of an MSO, as $\mathbf{MSO}$.

From each MSO we can extract $|\mathcal{C}'|$ different residual generators, one for each $c_i \in \mathcal{C}'$. The rest of the equations in $\mathcal{C}'$ form a just-constrained system for which a perfect matching can be found. This matching provides a calculation method for all the variables used by that residual generator. It should be noted that the number of possible MSOs grows exponentially large with the number of redundant equations in the initial system \cite{Krysander2005}.

In order to reduce the number of candidate MSOs, it is beneficial to consider only MSOs with at least one equation which can fail. A residual generator which involves only constraints which cannot fail is not useful and clutters the residual selection procedure \cite{Krysander2010}.

%% file: causality_calculability.tex
\section[cc]{Causality and Calulability}  \label{sec: cc}

The previously mentioned structural methods, regarding the selection of residual generators in large scale systems, provide only best-case results.
%In the previous section, a number of structural methodologies have been presented for the extraction of residual generators in large scale systems. However, it should be emphasized that the aforementioned structural methods provide only best-case results.
%In the process of creating the structural graph of the model, crucial information is lost and, consequently, results from standard structural analysis methods may prove to be unusable in the implementation phase of the FDI system. Rephrasing, \textit{a posteriori} examination of the extracted residual generators might show they are in fact unusable or at least, non-optimal according to certain criteria.
During the creation of the structural graph crucial information is lost and structural analysis may yield residual generators which are unusable or at least, non-optimal according to certain criteria.

Factors which can affect the feasibility and quality of the extracted residual generators are presented in this section. Namely, we will focus on the topics of \textit{causality}, \textit{calculability} and \textit{calculation cost}.
\vspace{-1pt}

\subsection{Causality}
Given an analytical equation,
we might deny the ability to solve it for a specific variable. Such a decision might stem from poor sensitivity, the existence of multiple solutions or simply the absence of an inverse function.
%As an example, it may be undesirable to solve any of the following three equations for $x$: in (\ref{eq: poorsens}) small measurement noise in $y$ would result in large variations in the solution, in (\ref{eq: trig}) multiple solutions are possible and in (\ref{eq: height}), a height map $h$ with arguments $x$ and $y$, no method to extract the correct $x$ value exists.
%\begin{IEEEeqnarray}{rCl}
%	y & = & \num{5d-5} x \label{eq: poorsens} \\
%	y & = & \cos(x) \label{eq: trig}\\
%	h & = & h(x,y) \label{eq: height}
%\end{IEEEeqnarray}

The existence of dynamic equations raises another concern, regarding the ability of the actual, automated FDI system to perform either numerical differentiation or integration \cite{Flaugergues2009, Blanke2006}. If our FDI system has the ability to perform numerical differentiation but not integration, then it is said that it operates under \textit{differential causality}. If the inverse is true,
%i.e. our system has the ability to perform numerical integration but not differentiation,
e.g. due to unacceptable noise amplification, it is said that it operates under \textit{integral causality}.
The combination of the above, which allows for both differentiations and integrations is called \textit{mixed causality} \cite{Svard2010}.
Differential and integral causalities pose restrictions in the admissible structural graph edges which can be incorporated in a matching set. 
%Under differential causality, an edge corresponding to integration cannot belong to a matching. Under integral causality, an edge corresponding to differentiation cannot belong to a matching.

%In our UAV example, the maximum matching that was found does obey differential causality: none of the integrated variables are matched by the differential constraints $d_1$ through $d_5$. On the other hand, the example matching does not obey integral causality: $\dot{q}$ and $\dot{h}$ are matched matched to the differential constraints $d_3$ and $d_5$ respectively. Thus, our matching obeys mixed causality.

Causality restrictions can be applied and visualized by adding directionality in the related edges of the graph, resulting in a partially-directed causal bipartite graph $\mathbf{G}_C {=} \left\{ \mathcal{C}, \mathcal{X}, \mathcal{E}_C \right\}$. Yet, this poses a serious limitation, since most computationally cheap matching tools are applicable only to undirected bipartite graphs.

\subsection{Calculability}\label{calculability}

The available computational solution tools also pose a constraint to the admissible matchings. Differentiators and integrators have already been mentioned in the previous subsection, but more types exist.
The DM decomposition of any just-constrained system may yield K\"{o}nig-Hall components of size larger than 1, as seen in Fig. \ref{fig: dm}, denoted by $\mathbf{G}^0_i$. Any matching spanning those components will produce a directed graph $\mathbf{G}_d$ with cycles, also known as Strongly Connected Components (SCCs).
Since equations belonging in the same cycle have to be solved simultaneously, simultaneous equation solver tools must be available  \cite{Svard2010, Katsillis1997}.
If a K\"{o}nig-Hall component contains only linear equations, then a Linear Algebraic Equation solver is required.
If it contains at least one non-linear algebraic equation, then a Nonlinear Algebraic Equation solver is required.
If it contains at least one dynamic equation, then a Differential Equation solver is required.
Thus, depending on the available tool set $\mathcal{T}$ in our FDI system, some matchings may be rejected.

\subsection{Additional Issues}

Even under differential (or mixed) causality, differential edges which belong in a loop (i.e. cycle) cannot be part of a calculable matching: Differential Equation solvers propagate the state variables by integration, not differentiation \cite{Flaugergues2010}.
Also, even under mixed causality, for practical applications, it is not desirable to add integral edges in a matching when they do not belong in a K\"{o}nig-Hall component, i.e. they are edges in a path of $\mathbf{G}_d$; noise bias will quickly invalidate the integration result.
We propose the term \textit{calculation causality} to capture the above restrictions.

\begin{figure}
	\centering
	\resizebox{0.45\linewidth}{!}{
		\input{figures/ex_rankcomp.tex}
	}
	\caption{A complete matching on variables in a cyclic graph, using the Ranking algoritm}
	\label{fig: ex_rankcomp}
\end{figure}
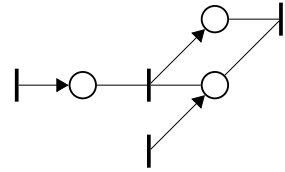

The final remark concerns the Ranking matching algorithm, which is a lucrative algorithm thanks to its low computational cost. As mentioned, the Ranking method is not guaranteed to find a complete matching on $\mathcal{X}_U$ in cyclic graphs. However, it is worth considering since there are cases where it can produce complete matchings, as seen in Fig. \ref{fig: ex_rankcomp}.
\vspace{-12pt}

%% file: figures/ex_rankcomp.tex
\begin{tikzpicture}

%\usetikzlibrary{arrows.meta}

\tikzstyle{every node}=[font=\small]
\tikzstyle{var}=[circle, draw=black, minimum width = 4mm, thick]
\tikzstyle{constr}=[rectangle, fill=black, draw=black, ultra thick, inner sep=0pt, minimum width = 0pt, minimum height = 5mm]
\begin{scope}[ node distance = 10mm]

\node [constr]	at (0,0) (c1)	[]		{};
\node [var]		(x1)	[right of = c1]	{};
\node [constr]	(c2)	[right of = x1] {};
\node [var]		(x3)	[right of = c2] {};
\node [var]		(x2)	[above of = x3] {};
\node [constr]	(c3)	[below of = c2] {};
\node [constr]	(c4)	[right of = x2] {};

\end{scope}

\begin{scope}[>=triangle 60]

\path[-]

(c1) edge [->]	(x1)
(x1) edge [-]	(c2)
(c3) edge [->]	(x3)
(c2) edge [-]	(x3)
(c2) edge [->]	(x2)
(x2) edge [-]	(c4)
(x3) edge [-]	(c4);

\end{scope}

\end{tikzpicture}

%% file: weights.tex
\section{A Weighted Graph Approach} \label{sec: we}

As was previously mentioned, matching a variable with a constraint implies solving the constraint for that variable. It may be easier, from a computational standpoint, to solve one constraint for one of its variables than for another. Encoding this kind of information in the structural model in the form of weighted edges provides a compact and easily manipulable data structure,
%In this section, the benefits of such a representation will be laid out.
which can then be used to discourage matchings of high computational cost and shorten the iteration time of a real-time FDI system.
%It is of interest to discourage matchings of high computational cost in the premise of shortening the iteration time of a real-time FDI system. In that spirit, a cost can be assigned to each way a constraint can be solved.
\input{simpleCosts.tex}
The actual costing information may be derived from computer science resources, engineering expertise or symbolic mathematics software.

Commonly in computing systems, addition and subtraction are of equal computational cost, multiplication is less ten times more expensive, division is about ten times more expensive and trigonometric functions and powers are almost 100 times more expensive. Square root cost varies with implementation but is generally expensive.

%As an example, consider constraint (\ref{eq: solveExample}), which is cheaper to solve in its original form, rather than be solved for $\delta_e$:
%\begin{equation}
%\delta_e = (\dot{q} - M_uu - M_ww - M_qq)/M_{\delta_e}
%\end{equation}

Regarding causality, edges which imply integration can be assigned greater cost, since integration requires storage space to hold the integrator state. Differentiation may be assigned an even greater cost, since it has the undesired effect of noise amplification.

Finally, information on sensor noise can be inserted in the structural graph by weighting corresponding edges proportionally to the noise variance of the corresponding sensor discouraging the use of a high-noise sensor.

%% file: simpleCosts.tex
The costing information can be encoded in the structural graph in the form of edge weights. Let an edge $e_i{=}(c_i,x_j) \in \mathcal{E}$ be assigned with a weight $w_m(e_i)$, via the weight function $w_m(\cdot)$, representing the cost of solving constraint $c_i$ for variable $x_j$.

Let there be a graph $\mathbf{G}{=}\left\{\mathcal{C},\mathcal{X},\mathcal{E} \right\}$ with a perfect matching $\mathcal{M}$, a weight function $w_m(\cdot)$ and a variable $x \in \mathcal{X}$.

\begin{definition}[Matching Cost]
	The cost of the matching is
	\begin{equation}
	J(\mathcal{M}) = \sum w_m(m_i), m_i \in \mathcal{M}
	\end{equation}
\end{definition}

\begin{definition}[Residual Generation Cost]
	Let there be a residual generator $c_i$, the matching-induced graph $\mathbf{G}_d$ and its inverse $\mathbf{G}_d'$.
	Let the reachable variable set from $c_i$ in $\mathbf{G}_d'$ be $\mathcal{X}_r$. Take the subset $\mathcal{M}_r \subseteq \mathcal{M} : \mathcal{M}_r {=} \left\{ (c_k, x_l): x_l \in \mathcal{X}_r \right\}$. The cost of the residual calculation is:
	\begin{equation}
	J(c_i) = \sum w_m(m_j) + w_r(c_i), m_j \in \mathcal{M}_r
	\end{equation}
	where $w_r$ is the evaluation cost of a residual generator from its variables.
\end{definition}

%It is worth noting that for any constraint $c_i$ and its variables $x_j \in var_U(c_i)$, it holds that
%\begin{equation}\label{eq:wrVSwm}
%w_r(c_i) = \underset{j}{min}\left( {w_m((c_i, x_j))} \right) + J_{-}
%\end{equation}
%
%\noindent where $J_-$ is the cost of a subtraction (see subsection \ref{operatorCost}). (\ref{eq:wrVSwm}) states that the residual generation cost is only marginally higher than the cheaper variable evaluation for $c_i$.

In \cite{Fravolini2009} a weight function was used to compare residual generators, but by pricing the operators in the analytical residual generator expression, not the structural graph edges.

A comparison among matching costs can be done in several levels:

\begin{enumerate}
	\item Each residual is assigned a residual generation cost, which constitutes a selection criterion for the composition of the FDI scheme.
	\item Within $\mathbf{MSO}_i$, for residual generator $c_j$, comparing matching costs for all feasible matchings $\mathcal{M}_k$ can lead to the selection of the cheapest calculation order. Let that cost be $J(\mathbf{MSO}_i,c_j)_{min} = \underset{k}{min} \left(J(\mathcal{M}_k)\right)$.
	\item Within $\mathbf{MSO}_i$, comparing the cheapest matching cost of each potential residual generator $c_j$ produces the cheapest residual generator for $\mathbf{MSO}_i$. Let that cost be $J(\mathbf{MSO}_i)_{min} {=} \underset{j}{min}\left( J(\mathbf{MSO}_i,c_j)_{min}\right)$.
\end{enumerate}

%% file: proposal.tex
\section[pr]{Proposed Residual Generation Method}  \label{sec: pr}

Existing methods perform post-processing of results to obtain maximal, calculable matchings of minimum cost in cyclic graphs. That requires the whole set of possible matchings be enumerated and then post-processed. This is problematic, since the number of different possible matchings grows exponentially with the number of vertices. Even by examining the over-constrained part only, the problem of choosing $|\mathcal{X}^+|$ out of $|\mathcal{C}^+|$ constraints to match variables of equal number can yield as many as 
${|\mathcal{C}^+|!}/{\left(|\mathcal{C}^+|-|\mathcal{X}^+| \right)!}$ matchings \cite{Krysander2008}.
MSOs help reduce the number of candidate matchings, but the pool can still remain significantly large.

We propose a method for shortening the time required to produce the set of candidate residual generators and select the optimal ones, through a combination of a priori and a posteriori graph processing.

\subsection{Step 1: DM Decomposition}
Conventionally, a DM decomposition is applied on the undirected system graph to extract the over-constrainted part. This provides the best-case detectability possibilities.

\subsection{Step 2: A Priori Matching Propagation}
Next, a maximal, calculable, loop-less matching of minimum cost is sought. Weights are applied to graph edges, as was discussed in the previous section. For this step, differential causality is enforced to avoid open-loop integrations. The Weighted Elimination matching algorithm is introduced to find a matching with the aforementioned properties. This algorithm is similar to the Ranking algorithm, in that it matches a variable to a constraint only if all the other variables of that constraint are known.

\begin{algorithm}
\caption{The Weighted Elimination algorithm}
\begin{algorithmic}[1]
	\Procedure{WeightedElimination}{$\mathbf{G}(\mathcal{C},\mathcal{X},\mathcal{E},w_m)$}
	\State $\mathcal{M} = \emptyset$
	\State $\mathcal{M}^* = \left\{(c_i, x_j) : |var_U(c_i)|=1\right\}$ \label{stepA:initMatchPool}
	\While{$\mathcal{M}^* \neq \emptyset$}\label{stepA:check}
		\State $J(m_i) = w_m\left((c_i, x_j)\right), \forall m_i=(c_i, x_j) \in \mathcal{M}^*$\label{stepA:newCost}
		\State Pick $m_{min} = arg\: \underset{m_i}{min}\left( J(m_i) \right)$\label{stepA:findMin}
		\State $\mathcal{M} = \mathcal{M} \bigcup {m_{min}}$
		\State $\mathcal{X}_U = \mathcal{X}_U \setminus var(m_{min})$
		\State $\mathcal{M}^* = \mathcal{M}^* \setminus $
			\Statex \hskip\algorithmicindent\hskip\algorithmicindent\hskip\algorithmicindent$\left\{ \left\{ m_{min}\right\} \bigcup \left\{ (c_i,x_j): x_j = var(m_{min}) \right\}  \right\}$\label{stepA:mRemove}
		\State $\mathcal{M}^* = \mathcal{M}^* \bigcup \left\{ (c_i, x_j) : |var_U(c_i)|=1 \right\}$\label{stepA:mAdd}
	\EndWhile
	\State \textbf{return} $\mathcal{M}$
	\EndProcedure
\end{algorithmic}
\end{algorithm}

%The algorithm \textsc{WeightedElimination} works as follows.
$\mathcal{M}^*$ is the pool of candidate matching extensions $m_i$.
In line \ref{stepA:initMatchPool} $\mathcal{M}^*$ is initialized. While there exist candidate matching extensions (line \ref{stepA:check}), the cost contribution of each new matching extension is calculated (line \ref{stepA:newCost}). The matching extension with the smallest cost is selected (line \ref{stepA:findMin}) and added to the matching. The selected matching extension is removed from $\mathcal{M}^*$, along with other candidate extensions that would match the same variable (line \ref{stepA:mRemove}). New, feasible matching extensions are added to $\mathcal{M}^*$ and the loop repeats.
\todo[inline]{Investigate on the naming "elimination"}

After \textsc{WeightedElimination} is completed, some unmatched constraints with all of their variables matched may exist. These constitute residual generators and are added to the candidate residual generator pool.

\todo[inline]{Present the algorithm properties}

\subsection{Step 3: A-Posteriori Matching Selection}
Commonly, K\"{o}nig-Hall blocks will exist and \textsc{WeightedElimination} will not return a maximal matching. The variables which are matched so far are considered known and the remaining subgraph can now be searched for MSOs.

Each MSO is searched for the cheapest, calculable residual generator evaluation. In order to find the minimum cost matching efficiently, we implemented Murty's algorithm \cite{Murty1968} to obtain all possible matchings in order of increasing cost.

The algorithm \textsc{GetOptimalResidual} finds the cheapest, calculable residual generator expression which respects calculability constraints for a given $\mathbf{MSO}_i{=}\left\{\mathcal{C},\mathcal{X},\mathcal{E}\right\}$.

%\vspace{-7pt}
\begin{algorithm}
	\caption{Find the optimal residual of an MSO}
\begin{algorithmic}[1]
	\Procedure{GetOptimalResidual}{$\mathbf{MSO}_i$}
	\For{$c_j \in \mathcal{C}$} \label{stepB:cLoop}
		\State $\mathbf{G}_R = \left\{ \mathcal{C}\setminus\left\{c_j\right\}, \mathcal{X}, \mathcal{E}' \right\}$
		\For{$\mathcal{M}_k \in$ \textsc{Murty}($\mathbf{G}_R$)} \label{stepB:Murty}
			\If{\textsc{IsCalculable}($\mathbf{G}_R, \mathcal{M}_k,\mathcal{T}$)}
				\State $J(\mathbf{MSO}_i,c_j)_{min} = J(\mathbf{MSO}_i,c_j,\mathcal{M}_k)$ \label{stepB:assignMatch}
				\State \textbf{break}
			\EndIf
		\EndFor
		\State $\mathcal{M}_j = \mathcal{M}_k$
	\EndFor
	\State $J(\mathbf{MSO}_i)_{min} = \underset{j}{min} \left(J(\mathbf{MSO}_i, c_j)_{min}\right)$
	\State \textbf{return} $\left\{ c_j, M_j \right\}, j= arg \: \underset{j}{ min} \left(J(\mathbf{MSO}_i, c_j)_{min}\right)$ \label{stepB:chooseResidual}
	\EndProcedure
\end{algorithmic}
\end{algorithm}
%\vspace{-7pt}

The function \textsc{Murty} returns the set of all the perfect matchings of the provided just-constrained graph, sorted by increasing cost.
\textsc{IsCalculable} is given a just-constrained graph, a matching and a set of solver tools and answers whether this matching can be calculated based on those tools.

\textsc{GetOptimalResidual} is given an MSO set and for every candidate residual generator (line \ref{stepB:cLoop}) it uses \textsc{Murty} to obtain all perfect matchings in ascending cost. The first calculable matching is selected (line \ref{stepB:assignMatch}). Once all residual generators have been examined, the one with the smallest cost is selected for the input MSO (line \ref{stepB:chooseResidual}).

As matchings are examined in increasing cost order, if at least one calculable matching exists for the residual generator $c_j$, it is found in at most $|\mathcal{C}|{-}1$ iterations and the cheapest residual will be available in at most $|\mathcal{C}| \cdot \left( |\mathcal{C}| {-}1 \right)$ iterations, where $|\mathcal{C}|$ is the number of constraints in said MSO.

Each MSO is examined and one residual generator is extracted from it. Afterwards, residual generators are selected in ascending cost order, until the specified detectability or isolability criteria are met.

%% file: example.tex
\section[ex]{Application Example}  \label{sec: ex}

A dedicated Matlab class for graph representation and manipulation was created, preserving the partially directed graph structure instead of the biadjacency matrix representation, which cannot hold directionality information.

\subsection{The System Model}

We applied the proposed method on the mathematical model of a fixed-wing UAV, comprized of 122 equations and 162 variables. The equations describe a multitude of submodels, ranging from the standard rigid-body kinematics, to aerodynamics, to propeller dynamics, to electric motor dynamics, to sensor models.
In total, there are 26 equations denoted as $k(\cdot)$ for the rigid body kinematics, 48 equations $f(\cdot)$ for the rigid body dynamics, 13 equations $d(\cdot)$ state explicitly the differentiated states, 6 equations $m(\cdot)$ for the atmosphere model and 29 equations $s(\cdot)$ for the measured quantities. A full equation list wouldn't fit in this paper.

Out of the complete equation set, we allowed only 58 equations to be subject to faults. Namely, $k8-k11$, $f17-f30$, $f34-f37$, $f39-f41$, $f43$, $f44$, $f46$, $f48$, $s1-s29$.

We evaluated the addition and subtraction operations with a weight of 1 cost unit, multiplication with 5, division with 10, roots, powers, and trigonometric functions with 100, integration with 100 and differentiation with 200. The weight assigned to each edge is equal to the total cost of the corresponding mathematical expression.

We assumed that integrators, differentiators, linear, non-linear and differential equations systems solvers were available.

\subsection{Results}

\begin{figure}
	\centering
		\includegraphics[width=\linewidth]{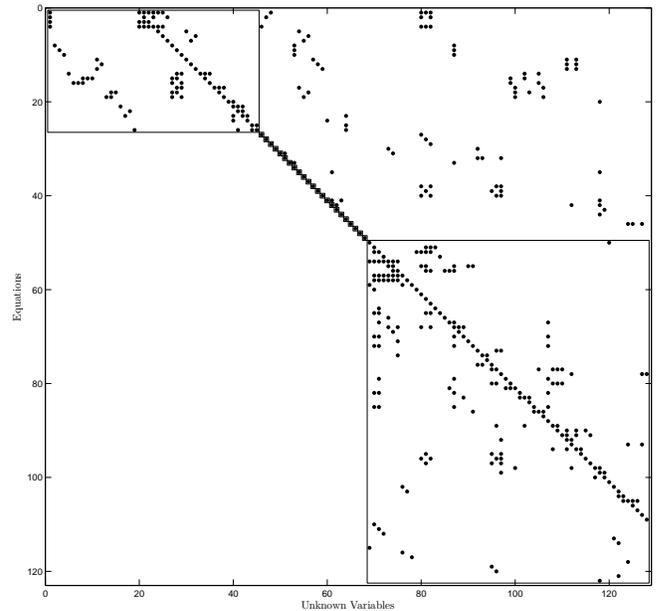}
	\caption{The DM decomposition of the example system}
	\label{fig: app_DM}
\end{figure}

Initially, DM decomposition of the undirected system graph was performed, as shown in Fig. \ref{fig: app_DM}.

26 equations and 45 unknown variables belong to the under-constrained part, 23 equations and unknown variables belong to the just-constrained part and 73 equations and 60 unknown variables belong to the over-constrained part. Thus, out of the possible 58 faults, only the 34 ones which correspond to $\mathcal{C}^+$ can be detected in the best case.

By applying the \textsc{WeightedElimination} algorithm to the resulting over-constrained part, we were able to match 42 variables and 50 equations. The 8 extra equations were assigned as candidate residual generators. The remaining graph is of much smaller size, and hence, lower complexity.

Then, we extracted all the MSO sets from the remaining graph. The analysis showed that out of 350 MSOs, 349 involved equations with potential faults. The cheapest, feasible residual generators of each MSO were stored, along with their corresponding cost.

The residual selection criterion was maximum detectability, but maximum isolability would be valid as well.
%The residual selection criterion was maximum detectability. Another valid option would be maximum isolability.
%By examining each residual generator in increasing cost order, the ones that extended the fault detectability of the FDI system were chosen.
In increasing cost order, the residual generators which extended fault detectability of the FID system were chosen.
Table \ref{fig: fsm} shows the resulting fault signatures matrix.

\begin{table}
	\centering
	\caption{The achieved detectability matrix}
	\includegraphics[width=\linewidth]{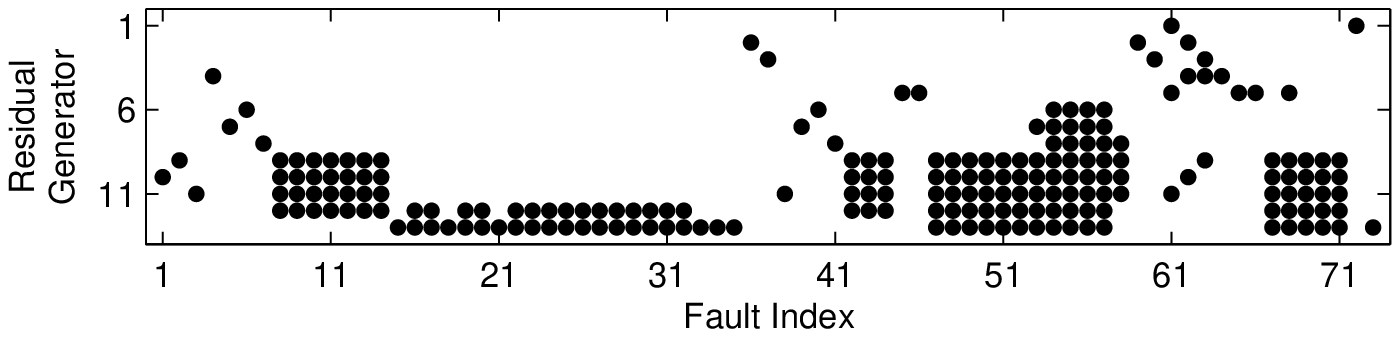}
	\label{fig: fsm}
\end{table}

In total, 13 residual generators were needed to achieve full fault detectability. Their costs are presented in Table \ref{fig: resList}.

\begin{table}
	\caption{The selected residuals and their cost}
	\input{residual_weights.tex}
	\label{fig: resList}
\end{table}

Two selected residuals are presented in detail. Residual generator $d1$ uses only 2 matching edges to evaluate the residual. The reachable subgraph from $d1$, involving equations $s13$ and $s16$ does not form any loops, hence no simultaneous equation solving tools are required to calculate the residual. This expected, since this residual generator was found during the application of \textsc{WeightedElimination}.

The affecting equations, involving part of a GPS sensor measurements are:
\vspace{-1pt}
\begin{IEEEeqnarray}{lCrCl}
	d1 & : & \dot{n} &=& ({d}/{dt})n\\
	s13& : & n_m & = & n \\
	s16& : & \dot{n}_m & = & \dot{n}
\end{IEEEeqnarray}
where $n$ is the Northwards position. This residual generator can be used to monitor the GPS sensor health.
%In detail, the residual generation function is, via direct substitution
\begin{IEEEeqnarray}{rCl}
r_{d1} &=& ({d}/{dt})n - \dot{n} = ({d}/{dt})n_m - \dot{n}_m
\end{IEEEeqnarray}

Another residual generator is $k1$. It was found during the MSO search phase and spans both the initial matching produced by \textsc{WeightedElimination} as well as the matching selected for its MSO. 28 equations are involved in its calculation: 8 as part of the MSO ($k12$, $k13$, $k14$, $d10$, $d11$, $s1$, $s2$, $s3$) and the rest as part of the initial matching ($s17$, $s10$, $s11$, $s12$, $k17$, $k19$, $s24$, $k20$, $s25$, $m5$, $s23$, $s21$, $m4$, $s22$, $k18$, $s4$, $s6$, $d12$, $s5$).

The constraints which where matched during \textsc{WeightedElimination} can be calculated by substitution. The remaining 8 equations
%which belong in a SCC, form the following system, which needs to be solved simultaneously.
need to be solved simultaneously.
\begin{IEEEeqnarray}{lCrCl}
	k12 & : & \dot{u} &= & rv -qw + {F_{bx}}/{m} \\
	k13 & : & \dot{v} &= & -ru + pw + {F_{by}}/{m} \\
	k14 & : & m & = & F_{bz} \ \left( \dot{w} - qu - pv \right)\\
	d10 & : & u &=& \begingroup\textstyle\int\endgroup \dot{u} dt\\
	d11 & : & v &=&  \begingroup\textstyle\int\endgroup \dot{v} dt\\
	s1  & : & F_x &=& m\left(a_{mx}  - g \sin \theta \right)\\
	s2  & : & F_y &=& m \left(a_{my} + \sin \phi \cos \theta \right)\\
	s3  & : & F_z &=& m \left(a_{mz} + \cos \phi \cos \theta \right)
\end{IEEEeqnarray}
$k12$-$k14$,$d10$ and $d11$ are part of the standard rigid-body kinematics equation set. $s1$-$s3$ are measurements from a 3-axis accelerometer, solved for the force component. This is a dynamic system which can be solved by a differential equations solver.

%% file: residual_weights.tex
\begin{small}

\tabcolsep=0.11cm
\setlength\extrarowheight{2pt}
\noindent
\begin{center}

\begin{tabularx}{\linewidth}{|c||*{7}{>{\centering\arraybackslash}X|}}
	\hline
	Generator & s26 & d1   & d2   & k4   & m2   & k6   & k5  \\ \hline
	  Cost    & 2   & 3    & 3    & 4    & 6    & 205  & 206 \\ \hline
	Generator & k7  & k2   & k1   & k3   & f12  & f12  &  \\ \hline
	  Cost    & 206 & 1356 & 1356 & 1556 & 2206 & 2218 &  \\ \hline
\end{tabularx}

\end{center}
\end{small}

%% file: conclusion.tex
\section{Conclusions}\label{sec: conclusion}

In this paper, SA methods were employed to perform fault diagnosis in large-scale systems.
Previous similar works were presented, which employed graph-theoretical methods to investigate the existence of analytical redundancy relations used as residual generators for fault diagnosis. Yet, those approaches did not take into account the implementation issues stemming from either causality restrictions, or calculability problems, or the existence of sets of simultaneous equations.

We introduced a method for finding causal, calculable residual generators of minimum cost, based on a novel, weighted bipartite graph representation. The combination of a priori and a posteriori information allows us to produce results in reduced time.

We applied the proposed method onto an example, large-scale model of a fixed-wing UAV and showed that different types of residual generators can be obtained, depending on the specified search criteria.